 \documentclass{emulateapj}
\def\spose#1{\hbox to 0pt{#1\hss}}
\def\lta{\mathrel{\spose{\lower 3pt\hbox{$\mathchar"218$}}
     \raise 2.0pt\hbox{$\mathchar"13C$}}}
\def\gta{\mathrel{\spose{\lower 3pt\hbox{$\mathchar"218$}}
     \raise 2.0pt\hbox{$\mathchar"13E$}}}

\shorttitle{scaling relations of dissipationless mergers}
\shortauthors{Aceves et al.}
\begin{document}
%%%%%%%%%%%%%%%%%%%%%%%%%%%%%%%%%%%%%%%%%%%%%%%%%
%\large

\title{Scaling Relations in Dissipationless Spiral-like Galaxy Mergers}  
\author{H. Aceves and H. Vel\'azquez}
  \affil{Instituto de Astronom\'{\i}a, Universidad Nacional Aut\'onoma de
  M\'exico. \\
  Apdo. Postal 877, Ensenada B.~C. 22800 M\'exico}
\email{aceves@astrosen.unam.mx}

\and
\author{F. Cruz}
\affil{Bah\'{\i}a San Quint\'{\i}n 211, Ensenada B.~C. 22860. M\'exico.}

%%%%%%%%%%%%%%%%%%%%%%%%%%%%%%%%%%%%%%%%%%%%%%%%%%%%%%%%%%%%%%%%%%%%%%%

%%%%%%%%
\begin{abstract} %\baselineskip=15pt
We determine both representations of the Fundamental Plane  [$R_{\rm e}
  \propto \sigma_0^a \langle I \rangle_{\rm e}^{-b} \rm{ and } R_{\rm e} 
  \propto (\sigma_0^2 \langle I\rangle_{\rm 
  e}^{-1})^\lambda$]  and the luminosity--effective phase space
density ($L \propto f_e^{-\gamma}$) scaling relation for $N$-body
  remnants of binary 
mergers of spiral-like galaxies.
The main set of merger simulations involves a mass-ratio of the
  progenitors in the range of about 1:1 to 1:5, harboring or not a
  bulge-like component, 
and are constructed using a cosmological motivated model.  Equal-mass
  mergers are also considered.
   Remnants lead to average values for the scaling indices of  $\langle a
   \rangle  \approx 1.6$,  $\langle b\rangle \approx 0.6$,  $\langle \lambda
   \rangle  \approx 0.7$, and   $\langle \gamma\rangle \approx 0.65$. These
   values are consistent with those of $K$--band observations
  (Mobasher et al. 1999) of ellipticals: 
   $\langle a \rangle  \approx 1.5$,  $\langle b\rangle \approx 0.8$,
   $\langle \lambda \rangle  \approx 0.7$, and   $\langle \gamma\rangle
   \approx 0.60$.  The $b$
  index is, however, not well reproduced. 
  This study does not allow us to establish a conclusive preference for models
  with or without a bulge as progenitors.
Our results indicate that
  the  $L$--$f_{\rm e}$ and FP scalings might be determined  to a large
    extent by dissipationless processes, a result that appears to be in
    contradiction to other dissipationless results. 

\end{abstract}


%%%%%%%%%%%%%%%%%%%%%%%%%%%%%%%%%%%%%%%%%%%%%%%%%%%%%%%%%%%%%%%%%%%%%%%%%
%%%%%%%%%%%%%%%%%%%%%%%%%%%%%%%%%%%%%%%%%%%%%%%%%%%%%%%%%%%%%%%%%%%%%%%%%
%\baselineskip=25pt

\section{Introduction}\label{sec:intro}
%%%%%%%%%%%%%%%%%%%%%%%%%%%%%%%%%%%%%%%%%%%%%%%%%%%%%%%%%%%%%%%%%%%%%%%%%
%%%%%%%%%%%%%%%%%%%%%%%%%%%%%%%%%%%%%%%%%%%%%%%%%%%%%%%%%%%%%%%%%%%%%%%%%

A large set of observational (e.g.~Schweizer 1998, Struck~2005,
 Rothberg \& Joseph 2004, 2006) and theoretical (e.g.
Barnes 1998, Burkert \& Naab 2003, Naab, Jesseit
\& Burkert~2006) evidence shows that elliptical galaxies are
 consistent with the 
picture of being formed by mergers of spiral galaxies (Toomre
  1977). However, some
 matters are still open to discussion, such as the problem of stellar ages and
 metallicities (e.g. Peebles~2002, Renzini~2006, Naab \& Ostriker~2007).

Observationally, one global scaling relation displayed by ellipticals is
 the Fundamental Plane  (FP; Djorgovski \& Davis 1987,
Dressler  et al.~1987), a linear-log relation among the effective radius
 $R_{\rm e}$, the  central stellar 1-dimensional velocity dispersion
$\sigma_0$, and the mean  effective 
 surface brightness $\langle I \rangle_{\rm e}$: 
 $R_{\rm e}  \propto \sigma_0^a \langle I \rangle_{\rm   e}^{-b}$.  
Values of $a$ and $b$ depend on 
 several factors among the most important are: (i) the observed
 wavelength, (ii)  
 the fitting method, (iii) the assumed underlying light  profile in 
 early-type galaxies and, (iv) the width and brightness of
 the magnitude range  of the sample [e.g. J{\o}rgensen, Franz 
 \&  Kjaergaard 1996, Pahre et al. 1998, Mobasher et.~al~1999 (MGZA99), 
 Bernardi et.~al. 2003 (B03), Jun \& 
 Im 2008, D'Onofrio et al. 2008, La Barbera et al. 2008, Nigoche-Netro et al. 
 2009]. A wide range of  values  $a\approx \,$(1-2)  and $b\approx\,$(0.5-1)
 can    be found in the  diverse cited works.

If structural and kinematical homology is assumed, as
 well as a constant
 mass-to-light ratio, direct application of the physical virial theorem to
 observational related quantities leads
 to a FP with  coefficients $(a,b)=(2,1)$.  The
     discrepancy of these values with the observed ones is usually
     termed as the "tilt" of the FP; see also  (\ref{eq:fplane2}).

The overall features of the observed FP  are reproduced by merger
studies [e.g.  Gonz\'alez-Garc\'{\i}a \& van Albada 2003, Aceves 
\& Vel\'azquez 2005 (AV05), Boylan-Kolchin et al. 2005, Robertson~et~al.~2006
 (R06),  Dekel \& Cox~2006],  but all show differences in 
  values obtained for the indices $a$ and $b$ depending  on the details of the
  initial conditions (e.g. the 
  presence or not of a bulge
  and/or a gas component) used in the simulations. It is likely
 that a combination of broken structural and dynamical homology and 
 stellar population effects, along with a variation of the central
 mass-to-light 
 ratio,  may lead to the explanation of the tilt 
 (e.g. Scodeggio et al.~1998, Trujillo, Burkert \& Bell~2004, AV05, Riciputi
 et al.~2005, R06, Bolton et al. 2007, Hyunsung
 \& Im  2008). 

The study of La Barbera et al. (2008) points in the
 direction that the tilt of the FP is not driven by stellar
 populations but from other processes like a homology breaking. They
 derive tight values of $a\approx 1.5$ and $b\approx 0.75$ from a
 large sample of early-type galaxies by combining  SDSS and UKIDSS
 data in the optical and NIR. 
However,   D'Onofrio et al. (2008) find larger variances of the
distributions of the FP coefficients as derived from a sample of three
larger surveys (WINGS, NFPS and SDSS) suggesting that a such wide
range of values for $a$ and $b$ seems to be in contradiction with
the idea of a universal FP relation and speculate that the FP
corresponds to a bent surface.

 The importance of dissipative processes in the merger hypothesis  was
 early recognized from estimates  
 of the central physical phase-space density of ellipticals, that
 turned out to be   higher than that of spirals (e.g.  
Ostriker 1980, Carlberg 1986, Gunn~1987, Lake~1989, Kormendy~1989). In a  
dissipationless scenario of galaxy merging, Liouville's theorem (e.g. Binney \&
Tremaine 1987) demands that
the central physical phase-space density remains constant so, given
the previous   observational estimates, it
prohibits a dissipationless picture of merging disk galaxies to form
 ellipticals. Robertson~et~al.~(2006) using a large set of 
   simulations  of equal-mass mergers without and with a gas component,
   and other physical processes,    conclude that the FP 
   in the $K$~band (Pahre et~al.~1998) and the effective radius -- stellar
   mass scaling relation [$R_{\rm e}$-$M_*$;  Shen~et~al. 2003 (S03)]  can be
   only reproduced by mergers of disk galaxies with a gas fraction  $f_{\rm
     gas} > 30$\%   with respect to the mass of the disk component, $M_{\rm
     d}$; ruling out dissipationless  mergers as mechanism to produce these
   two global scaling   relations.  Cox et~al. (2006) also find that $f_{\rm
     gas}$ is of the same order to reproduce the kinematic properties of
   ellipticals.

Obtaining higher central phase densities in models of elliptical galaxy
formation can be achieved, aside of taking into
account  dissipational processes, by including a
bulge-like component in the progenitors (e.g.  Vedel \& Sommer-Larsen~1990,
Hernquist, Spergel \& Heyl 1993; hereafter HSH).  
This result is in agreement with several $N$-body simulations 
 that  have found that mergers of pure disk galaxies, residing inside dark
 halos with a core-like profile,  are not able to reproduce the surface
 brightness density profiles typically found  
in ellipticals (e.g.
Garc\'{\i}a-Gonz\'alez \& Balcells 2005, Naab \& Trujillo 2006).

The combination of arguments based on the non-reproducibility of the FP,
the estimates of 
 phase-space densities and the density
profiles obtained from diverse $N$-body simulations have lead to the idea that
 collisionless models of
disk galaxies, with or without a bulge, 
 are ruled out as viable progenitors to form ellipticals
(e.g. Carlberg 1986, HSH, Mao \& Mo 1998, R06, O\~norbe et al.~2006, Robertson
 et al. 2006).     Nonetheless, there are results of $N$-body mergers of pure
 disk  galaxies residing in cuspy dark halos that lead to density  profiles 
 consistent with those of early-type  
 galaxies  (Aceves,  Vel\'azquez \&  Cruz 2006: AVC06) and are able to
 reproduce adequately the Fundamental Plane scaling relation (AV05). 
 Our results of dissipationless mergers (see below) suggest that  they can not
 be ruled out on the basis of scaling arguments.

On  the relevance of a bulge-like component in progenitors of
merger remnants,  there is observational 
 evidence that an important fraction of spirals do not harbor a bulge-like
 structure, or that it does not have a
 significant contribution, as is the case of  late type morphological
 Hubble types  
 usually called ``flat galaxies'' (e.g. Goad \& Roberts 1981, Karachentsev
 1989, Kautsch et~al. 2006, Dom\'{\i}nguez-Palmero et
 al. 2008). According to the sample considered by Kautsch  
 et~al. about one third of spirals are flat galaxies or simple disks,
 and in the sample of Dom\'{\i}nguez-Palmero et~al. about 30\% (54)
 harbor a bulge and about 70\% (137) have no  measurable bulge.
Thus, these flat galaxies arise the immediate issue 
about to what extent $N$-body simulations are able to establish  
global scaling relations like the FP when merging.

Undoubtedly, an important test on the details of the 
merger hypothesis  is imposed 
by the phase-space constraints. 
 A comparison among theoretical models of the central physical phase-space
density,  $f_{\rm p}$, using
different models and initial conditions, although important for our
understanding of the structure of remnants, needs to be related to
observations in order to determine their relevance. 
Different works have already shown that $f_{\rm p}$ increases with the
inclusion of a dissipative process (e.g. R06) or a bulge inside a core-like
halo (e.g. Naab \& Trujillo 2006). 
However, given that the
  physical  central phase-space density, $f_{\rm p}$, cannot be determined
  observationally an 
  estimator has to be used when comparing with observations.

An early estimator was the core
  coarse-grained phase-space density, $f_{\rm c}$, proposed by Carlberg
  (1986). However,  HSH  considered that 
  a more robust one is the  {\it effective} coarse-grained  
  phase-space  density, $f_{\rm e}$.
A strong dependency of $f_{\rm e}$ with the
luminosity, $L$, of ellipticals has been found in several works
(e.g. Carlberg 1986, Lake 1989, HSH, Mao \& Mo 1998).  For example,
using data of Bender, Burstein \& Faber (1992), HSH found    
that $L_B\propto f_{\rm e}^{-0.54}$ for the compact, intermediate and 
giant ellipticals with a {\sc rms} scatter of 0.15~mag; with $L_B$ being the
total luminosity in the $B$-band.

The FP and $L$--$f_{\rm e}$ scaling relations  appear to be tight enough
to serve as  constraints for the formation scenario of elliptical galaxies.
 However,  it should be said that up to now there is no direct
 comparison of the $L$--$f_{\rm  e}$  scaling   relation obtained 
 from observations with its counterpart provided by the remnants of  merger
 simulations.  Previous works  (e.g. HSH, R06)  have computed only 
 the physical cumulative coarse-grained distribution function, $s(f)$, 
 which is compared with a model of an elliptical galaxy or either
 with the effect of introducing a bulge and/or gas component in the
 progenitors, but it has not been related with an observational
 quantity.

By restricting the comparison of our numerical remnants with infrared 
observations, for which stellar population age effects are minimum
(e.g. Bruzual \& Charlot 2003),  
and by adopting a procedure similar to the one used 
in observational studies, we examine to what extent $N$-body numerical
simulations of spiral-like  galaxy mergers, with or without a bulge-like
component, are able to reproduce the observed   
$L$--$f_{\rm e}$, the FP and the $R$--$M_*$ global relations.
The latter two results are compared with those 
obtained in our earlier works (AV05, AVC06) where the progenitor 
galaxies lacked a bulge component and with recent results from
other authors. 

It is important to mention that we do not
attempt to determine the "zero-point" of the above scaling relations since
this surely requires to include more physical processes, such as star
formation, cooling, and feedback among others; a matter that  is out of
the scope of the present work.

This paper has been organized as follows.
 In Section~2 we summarize the
numerical galaxy models and tools used to carry out this work.
 Section~3 presents our results and  Section~4  contains a
 discussion and final comments.

\begin{table*}
\begin{center}
\caption{Properties of bulge progenitors of mergers.\label{tbl-1}}
\begin{tabular}{crccccrcrrrr}
	\hline
$Merger$ &  $M_d$        &    $R_d$   &  $z_d$  & $N_d$    &     $M_h$            & $r_{200}$ & $\lambda$ &  $c$  &  $N_h$     & $\theta$ & $\psi$  \\
         &  $(M_\odot)$ &   $(kpc)$  & $(kpc)$ &          &  $(M_\odot)$         &  $(kpc)$ &            &        &                     &           &           \\
	\hline
Mb01    & $4.82\times 10^9$ & $1.07$ & $0.17 $ & $32000$ & $5.42\times 10^{10}$ & $53.5$   & $0.053$    & $15.9$ &  $128000$ &  $0.05 $   & $ 5.93 $  \\
        & $3.44\times 10^9$ & $0.88$ & $0.13 $ & $22792$ & $6.01\times 10^{10}$ & $55.4$   & $0.03 $    & $12.1$ &  $142188$ & $1.43 $    & $5.29$    \\

Mb02    & $5.11\times 10^9$ & $2.34$ & $0.33 $ & $32000$ & $7.49\times 10^{10}$ & $59.6$   & $0.098$    & $7.9$  &  $160000$ & $0.01 $    & $4.39$     \\
        & $2.93\times 10^9$ & $1.77$ & $0.22 $ & $18340$ & $6.69\times 10^{10}$ & $57.4$   & $0.078$    & $13.3$ &  $142864$ &  $0.13 $    & $0.32$     \\

Mb03    & $3.31\times 10^9$ & $1.37$ & $0.23 $ & $20000$ & $4.88\times 10^{10}$ & $51.7$   & $0.074$    & $11.0$ &  $100000$ &  $2.87$    & $2.83$     \\
        & $8.25\times 10^9$ & $1.36$ & $0.26 $ & $49824$ & $9.21\times 10^{10}$ & $63.9$   & $0.054$    & $13.0$ &  $249112$ &  $1.95$    & $1.97$     \\

Mb04    & $1.31\times 10^9$ & $0.86$ & $0.17 $ & $20000$ & $1.02\times 10^{11}$ & $66.1$   & $0.023$    & $7.8$  &  $520000$ & $0.31$    & $0.69$     \\
        & $7.22\times 10^9$ & $4.41$ & $0.52$  &$110536$ & $8.11\times 10^{10}$ & $61.2$   & $0.099$    & $10.9$ &  $414487$ & $2.73$    & $5.47$     \\

Mb05    & $2.88\times 10^9$ & $1.55$ & $0.21 $ & $20000$ & $6.99\times 10^{10}$ & $58.3$   & $0.071$    & $9.9$  &  $160000$ & $1.97$    & $2.44$     \\
        & $9.14\times 10^9$ & $3.83$ & $0.56 $ & $63476$ & $1.27\times 10^{11}$ & $71.0$   & $0.109$    & $9.4$  &  $289635$ &  $1.41$    & $3.88$     \\

Mb06    & $7.41\times 10^9$ & $2.67$ & $0.45 $ & $32000$ & $8.76\times 10^{10}$ & $62.8$   & $0.088$    & $7.4$  &  $128000$ &  $1.74$    & $5.40$     \\
        & $4.63\times 10^9$ & $2.43$ & $0.46 $ & $19988$ & $5.64\times 10^{10}$ & $54.3$   & $0.076$    & $7.2$  &  $82508$  & $1.14$    & $5.92$     \\

Mb07    &$3.41\times 10^{10}$&$2.97$ & $0.49 $ &$120000$ & $4.18\times 10^{11}$ & $105.7$  & $0.076$    & $9.2$  &  $480000$ & $1.53$    & $2.08$     \\
        & $8.77\times 10^9$ & $1.02$ & $0.11 $ & $30868$ & $1.62\times 10^{11}$ & $77.1$   & $0.041$    & $7.9$  &  $186189$ &  $2.63$    & $1.16$     \\

Mb08    & $4.82\times 10^9$ & $1.70$ & $0.21$ & $48000$  & $1.01\times 10^{11}$ & $65.8$   & $0.046$    & $12.9$ &  $336000$ & $1.83$    & $4.27$     \\
        & $1.90\times 10^9$ & $0.72$ & $0.12$ & $18944$  & $5.70\times 10^{10}$ & $54.4$   & $0.038$    & $14.1$ &  $190491$ & $2.30$    & $3.98$     \\

Mb09    & $3.40\times 10^9$ & $5.37$ & $0.08$ & $40000$  & $1.28\times 10^{11}$ & $71.2$   & $0.021$    & $9.9$  &  $320000$ & $2.02$    & $5.54$     \\
        & $1.77\times 10^9$ & $1.01$ & $0.17$ & $20480$  & $8.50\times 10^{10}$ & $62.2$   & $0.037$    & $9.8$  &  $213107$ &  $1.34$    & $0.88$     \\

Mb10    & $3.43\times 10^9$ & $1.62$ & $0.16$ & $32000$  & $7.46\times 10^{10}$ & $59.5$   & $0.086$    & $13.3$ &  $256000$ & $2.02$    & $0.41$     \\
        & $6.19\times 10^9$ & $2.84$ & $0.33$ & $57832$  & $6.66\times 10^{10}$ & $57.3$   & $0.119$    & $14.0$ &  $228448$ & $0.26$    & $1.96$     \\

Mb11    &$5.61\times 10^{10}$& $3.46$& $0.36$ & $32000$ & $6.13\times 10^{11}$ & $120.2$   & $0.059$    & $6.3$  &  $128000$ & $2.61$    & $0.74$     \\
        &$4.34\times 10^{10}$& $5.34$& $0.74$ & $24776$ & $4.39\times 10^{11}$ & $107.5$   & $0.079$    & $7.3$  &  $91728$  &  $1.88$    & $6.02$     \\

Mb12    &$6.87\times 10^{10}$& $6.44$& $0.70$ & $60088$ & $1.27\times 10^{12}$ & $153.2$   & $0.070$    & $6.5$  &  $224245$ & $1.21$    & $1.88$     \\
        &$3.66\times 10^{10}$& $4.15$& $0.81$ & $32000$ & $1.45\times 10^{12}$ & $160.1$   & $0.045$    & $6.8$  &  $256000$ & $2.00$   & $3.33$     \\

Mb13    &$2.37\times 10^{10}$& $2.54$& $0.43$ & $32000$ & $3.65\times 10^{11}$ & $101.1$   & $0.057$    & $8.8$  &  $192000$ &  $0.46$    & $2.43$     \\
        &$4.68\times 10^{10}$& $2.86$& $0.42$ & $63156$ & $5.10\times 10^{11}$ & $113.1$   & $0.042$    & $7.8$  &  $268475$ &  $0.70$    & $6.19$     \\

Mb14    &$2.11\times 10^{10}$& $4.13$& $0.71$ & $32000$ & $3.60\times 10^{11}$ & $100.6$   & $0.050$    & $6.6$  &  $192000$ &  $0.98$    & $3.39$     \\
        &$2.14\times 10^{10}$& $1.07$& $0.16$ & $32404$ & $3.96\times 10^{11}$ & $103.9$   & $0.036$    & $14.2$ &  $211367$ &  $0.54$    & $5.33$     \\

Mb15    &$4.85\times 10^{10}$& $2.05$& $0.36$ & $60000$ & $8.54\times 10^{11}$ & $134.2$   & $0.049$    & $7.1$  &  $360000$ &  $1.72$    & $2.61$     \\
        &$2.11\times 10^{10}$& $4.61$& $0.55$ & $26052$ & $5.98\times 10^{11}$ & $119.2$   & $0.046$    & $6.8$  &  $251979$ &  $1.59$    & $0.43$     \\
\hline
\end{tabular}
\end{center}
$M_d$, $R_d$, $z_d$ and $N_d$ correspond to the mass, radial
  scale length, vertical scale length and number of particles for the disk
  component. $M_h$, $r_{200}$, $\lambda$, $c$ and $N_h$ indicate the mass,
  virial radius, spin parameter and concentration of the progenitor halos. The
  last two columns refers to Euler's random angles (in rads) used to rotate the
  axi-symmetric galaxy. In all cases, the bulge component, represented
  by a Hernquist 
  profile, was constrained to have a mass, scale length and number of
  particles of $M_d/4$, $R_d/3$ and $N_b/4$, respectively.

\end{table*}

%%%%%%%%%%%%%%%%%%%%%%%%%%%%%%%%%%%%%%%%%%%%%%%%%%%%%%%%%%%%%%%%%%%%%%%%%
%%%%%%%%%%%%%%%%%%%%%%%%%%%%%%%%%%%%%%%%%%%%%%%%%%%%%%%%%%%%%%%%%%%%%%%%%
\section{Simulations}
%%%%%%%%%%%%%%%%%%%%%%%%%%%%%%%%%%%%%%%%%%%%%%%%%%%%%%%%%%%%%%%%%%%%%%%%%
%%%%%%%%%%%%%%%%%%%%%%%%%%%%%%%%%%%%%%%%%%%%%%%%%%%%%%%%%%%%%%%%%%%%%%%%%

The method to build up our galaxy models is described in AV05 and for the sake 
of clarity it is briefly summarized. The galaxy models considered here
satisfy the Tully-Fisher relation 
(Tully \& Fisher 1977). They are set up by combining the Press-Schechter 
formalism of hierarchical clustering and  
 the cosmologically motivated model of disk galaxy formation of Mo, Mao
\& White (MMW; 1998). Essentially the method outlined by Shen, Mo \& Shu
(2002) is followed. 
In this scenario a formation redshift for disks needs to be
established, which was set here to be $z=1$ corresponding to a look-back time
of $\approx 8\,$Gyr (Peebles 1993); simulations with disks
formed at higher $z$ were not considered. In the MMW model, disks 
at higher redshifts are smaller and denser than those at the present
cosmic epoch. 
  A $\Lambda$CDM cosmological model with $\Omega_m=0.3$,
$\Omega_\Lambda=0.7$ and a Hubble parameter $h=0.7$ is adopted. 
The numerical progenitors satisfy the disk stability criteria of
 Efsthatiou, Lake \& Negroponte (1982),   $v_m^2/(GR_d^{-1}M_d) > 1.0$
 where $v_m$, $R_d$ and $M_d$ are the maximum rotational velocity, the radial
 scale    length and the total mass   of the disk.

The $N$-body galaxy models consist of a spherical dark halo, a stellar
disk and a bulge component.  
  The dark halo follows a Navarro, Frenk \& White (1997) profile
 with an exponential cutoff. The disk component has an
exponentially decaying radial density profile and an isothermal vertical
structure. Finally, the bulge-like component is represented by a
spherical Hernquist profile (Hernquist 1990). The bulge in this galaxy
formation scenario is introduced by following the procedure described  
by Springel \& White (1999).  In all cases, a fixed mass fraction of
$0.25 M_{\rm d}$   and a scale length of one-third of the radial
scale of the disk are adopted for the bulge. Table 1 summarizes 
the parameters of our progenitor galaxies with a central bulge component.

Also, for comparison purposes the remnants of our pure disk galaxy
models described in AV05 together with five new  similar merger
simulations are 
included. In this case, the progenitors of our remnants resemble the
"flat galaxies" studied by Kautsch et al. (2006) and
Dom\'{\i}nguez-Palmero et al. (2008). 

 Only parabolic encounters are considered
 with pericenter parameters ($5$--$20\,$kpc) consistent with those found in
cosmological simulations (e.g.~Navarro, Frenk \& White
1995). The spin orientation of each galaxy,
relative to the orbital plane and defined here using Euler's angles 
(Goldstein~1950), has been taken 
randomly. We have only  
considered binary encounters, so our results may not apply to massive
early-type galaxies since they are probably the result of multiple  disk galaxy
merger events   and/or mergers of early-type galaxies 
(e.g.  Weil \& Hernquist 1996; Naab, Khochfar \& Burkert 2006).

%%%%%%%%%%%%%

\begin{table*}
\begin{center}
\caption{Remnants properties with bulge progenitors.\label{tbl-mb}}
\begin{tabular}{ccccccccc}
	\hline
Merger & $n$     &  $R_{\rm e}$       &    $\langle I \rangle_{\rm e}$
&   $M_{\rm L}$   &   $\sigma_0$    &  $R_{50}$     &   $\sigma_{50}$
&  $M_{\rm L,50}$ \\

       &         &  $(kpc)$           &  $(M_\odot/kpc^2)$         &
       $(M_\odot)$     & $(km/s)$  & $(kpc)$  & $(km/s)$ & $(M_\odot)$
       \\ 
 
\hline

Mb01 & 3.48  & 1.59E+00 & -2.21E+01  & -2.51E+01  & 7.15E+01  & 1.41E+00 &  7.52E+01 & -2.43E+01  \\
     &  & 1.62E+00 & -2.20E+01  & -2.51E+01  &  7.15E+01   &  &  & \\

Mb02 & 3.08  & 3.23E+00 & -2.05E+01 & -2.50E+01  & 6.49E+01  &  2.61E+00 & 6.79E+01  & -2.43E+01 \\
     &  & 3.55E+00 & -2.03E+01  & -2.50E+01  & 6.51E+01    &  &  & \\

Mb03 & 3.23  & 1.95E+00 & -2.20E+01 & -2.54E+01  & 7.92E+01  &  1.45E+00 & 8.54E+01  & -2.46E+01 \\
     &  & 2.02E+00 & -2.19E+01 & -2.54E+01  & 7.94E+01  &  &  & \\

Mb04 & 2.89  & 4.41E+00 & -1.98E+01 & -2.50E+01  & 5.91E+01  &  3.82E+00 & 6.74E+01  & -2.43E+01\\
     &  & 4.77E+00 & -1.96E+01 & -2.49E+01  & 5.97E+01  &  &  & \\

Mb05 & 2.70 & 4.11E+00 & -2.04E+01 & -2.55E+01  & 7.31E+01  &  3.55E+00 & 7.65E+01  & -2.47E+01\\
     & & 5.29E+00 & -1.99E+01 & -2.55E+01  & 7.48E+01  &  &  & \\

Mb06 & 3.43  & 3.43E+00 & -2.08E+01 & -2.55E+01  & 6.90E+01  &  2.75E+00 & 7.31E+01  & -2.47E+01 \\
     &  & 3.69E+00 & -2.06E+01 & -2.54E+01  & 6.94E+01  &  &  & \\

Mb07 & 2.71  & 3.91E+00 & -2.19E+01 & -2.68E+01  & 1.04E+02  &  3.54E+00 & 1.08E+02  & -2.61E+01\\
     &  & 5.08E+00 & -2.14E+01 & -2.69E+01  & 1.07E+02  &  &  & \\

Mb08 & 2.42  & 2.00E+00 & -2.13E+01 & -2.48E+01  & 6.14E+01  &  1.97E+00 & 6.78E+01 & -2.41E+01 \\
     &  & 2.28E+00 & -2.11E+01 & -2.48E+01  & 6.21E+01  &  &  & \\

Mb09 & 3.59  & 1.21E+00 & -2.22E+01 & -2.46E+01  & 5.47E+01  &  1.20E+00 & 6.54E+01 & -2.38E+01 \\
     &  & 1.18E+00 & -2.22E+01 & -2.45E+01  & 5.45E+01  &  &  & \\

Mb10 & 3.11  & 3.28E+00 & -2.06E+01 & -2.52E+01  & 7.23E+01  &  2.32E+00 & 7.78E+01 & -2.44E+01\\
     &  & 3.78E+00 & -2.03E+01 & -2.52E+01  & 7.31E+01  &  &  & \\

Mb11 & 3.17  & 7.29E+00 & -2.14E+01 & -2.77E+01  & 1.38E+02  &  7.04E+00 &  1.33E+02 & -2.70E+01\\
     &  & 7.66E+00 & -2.12E+01 & -2.76E+01  & 1.38E+02  &  &  & \\

Mb12 & 2.58  & 8.55E+00 & -2.12E+01 & -2.78E+01  & 1.54E+02  &  7.90E+00 &  1.61E+02 & -2.70E+01\\
      & & 1.10E+01 & -2.05E+01 & -2.77E+01  & 1.56E+02  &  &  & \\

Mb13 & 2.86  & 5.14E+00 & -2.18E+01 & -2.74E+01  & 1.30E+02  &  5.07E+00 & 1.31E+02  & -2.66E+01 \\
      & & 6.37E+00 & -2.14E+01 & -2.74E+01  & 1.32E+02  &  &  & \\

Mb14 & 2.19  & 3.04E+00 & -2.32E+01 & -2.76E+01  & 1.98E+02  &  3.68E+00 & 1.78E+02  & -2.70E+01\\
      & & 6.01E+00 & -2.21E+01 & -2.80E+01  & 1.94E+02  &  &  & \\

Mb15 & 4.02  & 3.95E+00 & -2.24E+01 & -2.74E+01  & 1.43E+02  &  4.07E+00 & 1.38E+02  & -2.66E+01 \\
     &  & 3.98E+00 & -2.24E+01 & -2.73E+01  & 1.43E+02  &  &  & \\
\hline
\end{tabular}
\end{center}
First line of each merger corresponds to a S\'ersic fit, while the
second to a $R^{1/4}$. Simbols $n$, $R_{\rm e}$,   $\langle I \rangle_{\rm e}$, $M_{\rm L}$, and
$\sigma_0$ correspond, respectively, to the S\'ersic index, effective radius, mean
effective surface brightness, total ``magnitudes'' ($-2.5\log {\rm
  M}_{\rm L}$), and
central velocity dispersion within $R_{\rm e}/8$.  Values $R_{50}$,
$\sigma_{50}$ and $M_{\rm L,50}$ are the proyected half-luminous mass
radius, the 1D velocity dispersion, and the half-mass proyected
luminous mass (in ``magnitudes''), respectively, of our remnants measured directly from
the simulations. Units are indicated.
\end{table*}
%%%%%%%%%%%%%

 The progenitor galaxies of our  main set of merger simulations  have a
mass-ratio in the range of about 1:1 to 1:5. However, we have also added
  a new set of ten equal-mass mergers to assess the importance 
  of the progenitors masses in dissipationless 
mergers of disk galaxies in determining the FP and the $L$--$f_e$
global scaling 
relations. These simulations were obtained using as 
  progenitors galaxies the ones partaking in mergers $M01$, $M05$, $M07$,
  $M08$ and $M10$ of AV05 that lead to luminous-dominated merger cores. New
  orbital  parameters, following the above procedure, were obtained.

Simulations were done using 
the parallel tree-based code {\sc Gadget-2}
 (Springel, Yoshida \& White 2001, Springel~2005). 
They were evolved for about $8\,$Gyr, 
a time similar to the
time spanned from $z=1$ to the present epoch and at which the remnants  had
reached  equilibrium. Softening for disk, bulge and halo particles 
were taken to be $\epsilon_d=\epsilon_b=35$ pc and $\epsilon_h=350$ pc,
respectively, with a parameter {\sl ErrTolForceAcc}$=0.0025$.
Energy conservation was better than 0.75\% in 
all simulations.

%%%%%%%%%%%%%%%%%%%%
\begin{table*}
\begin{center}
\caption{Scaling indices.\label{tbl-2}}
\begin{tabular}{l | r| rrrcc|c|c|c}
\tableline\tableline
       &   & & \multicolumn{3}{c}{$R \propto \sigma^a \langle I
  \rangle_{\rm e}^{-b}$} & & 
 $R \propto  M_*^\mu$     &$L \propto f_{\rm e}^{-\gamma}$ & {\small Fit}\\
\tableline
Sample &  Profile & & $a$ & $-b$ & $\langle \Delta^2\rangle^{1/2}$  &
$\lambda$ & $\mu$         & $\gamma$  & \\ 
\tableline
No Bulge & $S$  &  &$1.56\pm 0.21$  & $0.69\pm 0.08$  & 0.06  & $0.71\pm0.04$ &
$0.73\pm 0.14$ &  $0.50\pm 0.08$  &  {\sc lsq}  \\
          &   &  & $1.72\pm 0.66$  & $0.68\pm 0.13$  &   & $0.72\pm0.04$ &
$0.86\pm 0.32$ &  $0.51\pm 0.08$  &   {\sc ort} \\

      & $R^{1/4}$ &  &$1.76\pm 0.29$  & $0.67\pm 0.08$  & 0.09  &
$0.68\pm 0.06$&  $0.59\pm 0.16$  & $0.51 \pm 0.11$ &  {\sc lsq} \\
      &          &  &$1.88\pm 0.40$  & $0.66\pm 0.11$  &   & $0.73\pm0.08$ &
 $0.80\pm 0.26$  & $0.55 \pm 0.11$ &  {\sc ort} \\

\hline

Bulge & $S$   & &$1.48 \pm 0.09$  & $0.59\pm 0.06$  & 0.05  & $0.66\pm0.05$ &
$0.34\pm 0.09$  & $0.72\pm 0.12$  &  {\sc lsq} \\
      &       & &$1.54 \pm 0.15$  & $0.62\pm 0.08$  &       & $0.68\pm0.05$ &
$0.36\pm 0.10$  & $0.84\pm 0.14$  &  {\sc ort} \\

      & $R^{1/4}$ & &$1.44\pm 0.09$  & $0.54\pm0.03$  & 0.04  & $0.63\pm0.04$ &
 $0.40\pm 0.09$  &$0.66\pm 0.10$ &  {\sc lsq}\\
     &            & &$1.47\pm 0.09$  & $0.55\pm0.04$  &   & $0.65\pm0.08$ &
 $0.47\pm 0.66$  &$0.76\pm 0.14$ &  {\sc ort}\\
% -    & $R_{50}$ & &1.70  & 0.62  & 0.04  &  & $0.40$ & & $0.74$ \\
\hline

No Bulge eM & $S$   & &$1.54 \pm 0.14$  & $0.48\pm 0.09$  & 0.04  &
$0.68\pm0.07$ & 
$0.47\pm 0.07$  & $0.71\pm 0.09$  &  {\sc lsq} \\
      &       & &$1.60 \pm 0.17$  & $0.50\pm 0.10$  &       & $0.70\pm0.09$ &
$0.50\pm 0.14$  & $0.75\pm 0.12$  &  {\sc ort} \\

      & $R^{1/4}$ & &$1.52\pm 0.16$  & $0.49\pm0.03$  & 0.05  & $0.62\pm0.06$ &
 $0.49\pm 0.11$  &$0.63\pm 0.11$ &  {\sc lsq}\\

     &            & &$1.55\pm 0.15$  & $0.49\pm0.07$  &   & $0.65\pm0.08$ &
 $0.53\pm 0.16$  &$0.66\pm 0.13$ &  {\sc ort}\\

      & $R_{50}$ & &$1.78\pm 0.11$  & $0.51\pm0.04$  & 0.03  & $0.79\pm0.09$ &
 $0.45\pm 0.06$  &$0.77\pm 0.08$ &  {\sc lsq}\\
      &   & & $1.80\pm 0.12$  & $0.53\pm0.08$  &   & $0.83\pm0.12$ &
 $0.47\pm 0.07$  &$0.80\pm 0.09$ &  {\sc ort}\\
\hline

R06  ($N$-body) & $R_{50}$  &  &2.00  & 1.01  & 0.02  & $1.00\pm0.01$
& $0.45\pm0.03$   & &  {\sc lsq} \\
\hphantom{R06} (Full model)& $R_{50}$  & &1.55  & 0.82  & 0.06  & $0.79\pm0.01$&
$0.57\pm0.02$   &  &  {\sc lsq} \\

\hline
%D'Onofrio & $S$   & & $0.99\pm 0.15$ & $0.57\pm 0.07$  & $0.12$  &  &
%$0.73 \pm 0.07$ &  &$0.51\pm 0.04$  \\
%          &       & & $1.26\pm 0.31$ & $0.52\pm 0.10$  &         &  &
%$0.78 \pm 0.07$ &  &$0.52\pm 0.03$  \\
%BBF & $R^{1/4}$   & & 1.35 & 0.84  & 0.12    &  & $0.75 \pm 0.03$ &  &$0.53 \pm
% 0.02$ \\
%   &         & &  &   &    &  & $0.79 \pm 0.04$ &  &$0.54 \pm 0.02$ \\

MGAZ99 & $R^{1/4}$   & & $1.38\pm 0.08$ & $0.74 \pm 0.04$  & 0.07    &
$0.72\pm 0.03$ & $0.62\pm 0.05$   &$0.57 \pm 0.03$ &  {\sc lsq} \\
       &            & & $1.53\pm 0.11$ & $0.77 \pm 0.04$  &     & $0.74\pm0.03$
& $0.68\pm 0.07$   &$0.60 \pm 0.04$ &  {\sc ort}\\

PDdC98 &  $R^{1/4}$  & & $1.53\pm 0.08$ & $0.79 \pm 0.03$  &     &
  & $0.57\pm 0.01$   &  &   {\sc ort}\\
 {\footnotesize \quad Coma}  &  & & $1.33\pm 0.19$ & $0.76 \pm 0.08$  &     &
&    &  &  {\sc ort}\\

%JI08 & $R^{1/4}$   & & $1.42\pm 0.11$ & $0.81 \pm 0.05$  &     &
%&    & & {\sc bis} \\

LaB08 & $S$   & & $1.51\pm 0.04$ & $0.77 \pm 0.01$  & 0.07    &
  &  & &  {\sc lsq} \\
      &       & & $1.53\pm 0.04$ & $0.77 \pm 0.01$  & 0.06    &
  &  & &  {\sc ort}\\

SDSS  (B03,S03) & $R^{1/4}$, $S$  & &$1.20\pm0.04$  & $0.76\pm0.01$
& 0.05  & & 0.56   &  & {\sc lsq} \\
$\phantom{SDSS}$   & $R^{1/4}$ & &$1.51\pm0.05$  & $0.77\pm0.01$
&   & &  &    &   {\sc ort}\\
\tableline 
\end{tabular}
\vspace{0.0in}
\end{center}
\end{table*}
%%%%%%%%%%%%%%%%%%%%%%%

%%%%%%%%%%%%%%%%%%%%%%%%%%%%%%%%%%%%%%%%%%%%%%%%%%%%%%%%%%%%%%%%%%%%%%%%%
%%%%%%%%%%%%%%%%%%%%%%%%%%%%%%%%%%%%%%%%%%%%%%%%%%%%%%%%%%%%%%%%%%%%%%%%%
\section{Results}
%%%%%%%%%%%%%%%%%%%%%%%%%%%%%%%%%%%%%%%%%%%%%%%%%%%%%%%%%%%%%%%%%%%%%%%%%
%%%%%%%%%%%%%%%%%%%%%%%%%%%%%%%%%%%%%%%%%%%%%%%%%%%%%%%%%%%%%%%%%%%%%%%%%

In this section we present the FP, $R_{\rm e}$-$M_*$ and the $L$-$f_{\rm e}$
relations for our dissipationless mergers of spiral-like galaxies. 

We resort to a procedure similar to the one used in practice with the
observations in order to avoid, as far as possible, any kind of biases
in our results difficulting  their comparison with them.  
 $K$-band observations are better suited for our
 purposes, since they are 
 less affected by  stellar population effects. For this
 reason we have chosen the observational
 data of MGAZ99 which is publicly available. 

The results for  bulgeless progenitors are obtained by 
 considering only  those mergers of our sample (10 out of 15) that resulted in
  remnants  dominated  by   ``luminous'' matter inside  their effective radius
  ($R\lta R_{\rm e}  \approx 1$-$2\,$kpc) as is suggested by observations
  (e.g. Gerhard   et al.~2001,  Thomas et  al.~2007). However, the
 scaling indices do not change significantly if    remnants  centrally
 dominated by dark matter are included. 

Each remnant was ``observed'' along 100 random different lines-of-sight and 
fitted with a $R^{1/4}$ (de Vaucoleurs~1953) and a S\'ersic
  profile (S\'ersic 1968). The
  $R^{1/4}$-profile was adopted to compare directly with MGAZ99 while the
  S\'ersic profile is better suited to fit our numerical remnants and it is
  characterized by the following surface density:

\begin{equation}
\Sigma (R) = \Sigma_0 \exp[ -b (R/R_{\rm e})^{1/n}] \,,
\end{equation}
where $b=b(n)\!\approx\! 2n-1/3+4/(405n)$ (e.g.~Graham \& Driver~2005), 
$\Sigma_0$ is the central projected surface luminous mass density and $n$ is
the corresponding S\'ersic index.  
The radial interval of the fit is taken from our numerical resolution value
for the luminous matter ($2.8\epsilon_d=2.8\epsilon_b \approx 100$~pc)
to the outer radius enclosing 95\% of the 
projected luminous mass, determined directly by counting particles from
the simulations. 
The parameters $\Sigma_0$, $R_{\rm e}$ and $n$ of the profile were found by a
$\chi^2$-minimization using  the Levenberg-Marquardt method (e.g. Press
et~al. 1992).

The total luminous mass of a S\'ersic model is given by
\begin{equation}
{\rm M}_{\rm L} = \frac{2 \pi n}{ b^{2n}}
  \Gamma(2n)\, \Sigma_0 R_{\rm e}^2 \, ,
\end{equation}
were $\Gamma$ is the Gamma function. 
 For a crude estimate of the luminosity, $L$, of the numerical remnants we may
 choose a  constant stellar mass-to-light ratio, $\Upsilon_*$, leading to
 $L=\Upsilon_*^{-1} {\rm M}_{\rm L}$.  
 To estimate a total magnitude $M_K$ in the
 $K$-band we
 take as a fiducial values $\Upsilon_*=0.97$, 
in solar units, and $M_{\odot}=3.32$.

The central projected, luminosity weighted, velocity
dispersion, $\sigma_0$,  of the luminous particles was
computed inside a circular region of radius $R_{\rm e}/8$ for each remnant
projection,  as is usually done in observational studies (e.g. J{\o}rgensen et
al. 1996, MGAZ99, D'Onofrio et al. 2008). A value for the mean effective
surface brightness $\langle I \rangle_{\rm e} $  can be
obtained from the fitted parameters of the profile  and by assuming a
constant $\Upsilon_*$.     Results of different quantities for our set of simulations with a
bulge progenitors are shown in Table~2.

%%%%%%%%%%%%%%%%%%%%%%%%%%%%%%
\subsection{Fundamental Plane}
%%%%%%%%%%%%%%%%%%%%%%%%%%%%%%

A log-plane of the form:
\begin{equation}
\label{eq:fplane}
\log R_{\rm e} \propto a \log \sigma_0 + b \log \langle I \rangle_{\rm  e}
\end{equation}
was fitted to our remnants values by two standard
  methods (e.g. B03):  the direct least-square method ({\sc
  lsq})\footnote{\footnotesize Formulae (9)  and (14) in Robertson et
  al. (2006) for the direct least-square method display two
  misprints; compare them with Bernardi et al. (2003).} and an
  orthogonal plane ({\sc  ort}) procedure.
    The tilt $\lambda$ of
  the FP given by
\begin{equation}
\label{eq:fplane2}
\log R_{\rm e} \propto \lambda\, \log \, (\, \sigma_0^2  \langle I
\rangle_{\rm  e}^{-1} \, )\,, 
\end{equation}
was also determined using the two previous fitting methods. 
 Both the {\sc lsq} and {\sc ort} methods were tested against synthetic data, 
in 2-D and 3-D, with an excellent agreement between the synthetic and fitted 
parameters.

             Table~3 summarizes the main
     results for our simulations with and without a bulge component,  and
       using either a S\'ersic (S) or $R^{1/4}$ profile.
     Columns (3-5) provide the 
     indices $a$ and $b$ along with the corresponding {\sc rms}. Column (6)
     provides the tilt ($\lambda$)  of
     the FP, and Column (7) shows the
     index $\mu$ for the fitting of the  relation $R \propto  M_*^\mu$. Column
     (8) presents the index $\gamma$ of the luminosity-phase space relation
     (see below), and the last column (9) indicates the fitting method
     used.  We have included also results for a set of simulations of
       equal-mass mergers (eM) lacking a bulge component. Errors in
     the {\sc ort} fits were estimated by bootstrap, while those in the
     {\sc lsq} fits is the standard {\sc rms}.

     For comparison purposes, we  have
     included in Table~3 the following:
     (a) values quoted by Robertson et~al. (2006) for their 
     N-body  and full ($N$-body+gas) simulations.  The methodology
       followed by these authors is, however, different to the one adopted
       here and in observational studies.  Robertson et~al. measure
     directly 
     a half-mass stellar effective radius ($R_{50}$) inside which the
     1-dimensional dispersion velocity, $\sigma_{50}$, and a mean
     surface density  
     $I_e={\rm  M}_{\rm L,50} (< R_{50})/\pi R_{50}^2$ are 
     determined; a 3D log-plane (\ref{eq:fplane}) is then fitted by a {\sc
           lsq} method. 
     (b) Some fits from observational data are also
     included. Data from MGAZ99 for  Coma ellipticals are used to
     compute different scalings. We indicate Pahre et al. (1998; PDdC98)
         average fitted parameters for the FP for all the clusters in their
         sample as well as the corresponding ones to the Coma cluster. The FP
         in the K-band from La Barbera et al. (2008; LaB08) is also shown. 
     Despite the SDSS data are in the optical region,
     we include the FP and $R$-$M_*$ relations derived by 
Bernardi et al. (2003) and Shen et al. (2003) using a $R^{1/4}$-{\sc
  lsq}-{\sc ort} and a S\'ersic-{\sc ort} profile, respectively.

%%%%%%%%
\begin{figure}[!t]
\centering
\includegraphics[width=8.5cm]{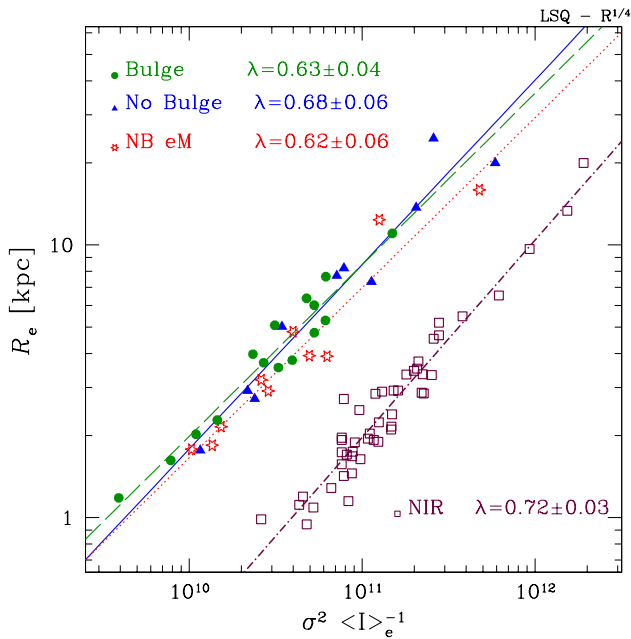}
\caption{Comparison of near-infrared (NIR) Fundamental Plane (Mobasher et
  al. 1999, {\it open squares}) 
  with our $N$-body remnants, with ({\it filled circles}) and without ({\it
    filled  triangles}) a bulge. Results for equal-mass 
  mergers, without a bulge (NB eM, {\it open stars}), are also shown. The
  ``tilt'' $\lambda$ 
  of the FP in each case is indicated: (a) mergers with ({\it long-dashed
    line}) and 
  without ({\it solid-line}) a bulge component, (b) equal-mass mergers without a
  bulge ({\it dotted line}) and, (c) K-band observation by MGAZ99 ({\it
    dot-dashed line}).} 
\label{fig:fplane}
\end{figure}
%%%%%%%%%%%%

 In Figure~\ref{fig:fplane} we plot the FP in the ``virial plane''
  representation:  $R_{\rm e}$--$\sigma_0^2
  \langle I\rangle_{\rm e}^{-1}$. Here, results from our
  $N$-body mergers, with ({\it solid dots})  and without ({\it filled
  triangles}) a bulge-like component, 
  are shown.  Equal-mass mergers lacking a
  bulge are also included ({\it open stars}). 
  For comparison, the FP in the K-band is represented by open
    squares and, finally, the tilt, $\lambda$, in each case is provided.

From Table~3, it follows that our $N$-body remnants show different 
values for the indices $a$, $b$, $\lambda$ and $\mu$, depending on
both the fitting
method and the assumed profile for the luminous matter. Averaging over
all the results, ignoring the details of the fitting, lead to $\langle
a\rangle \approx 1.6$, 
$\langle b\rangle \approx -0.6$, $\langle \lambda \rangle \approx 0.7$, and
$\langle \mu \rangle \approx 0.5$.  On the other hand,  the
corresponding $K$-band averages are:  $\langle a\rangle \approx 1.5$, 
$\langle b\rangle \approx -0.8$, $\langle \lambda \rangle \approx 0.7$, and
$\langle \mu \rangle \approx 0.6$.

The previous average values indicate a general  good agreement between the
theoretical and observational values; they are also consistent with
those obtained by Bolton~et~al.~(2007) using  lens galaxies:
$a=1.50\pm0.32$ and  $b=-0.78 \pm 0.13$.
However the index $b$, associated with the 
luminosity and not the stellar mass, is the one that shows a larger discrepancy
from the observed values. A similar thing occurs for the $\mu$ index.
Nonetheless the tilt, $\lambda$, obtained through the fit of the
combined variable  $\sigma_0/\langle I\rangle_{\rm  e}$, yields results that agree very
well with the observational data.  

Judging only from the individual indices
$a$ and $b$ it is not clear if models of progenitors with or without
a primordial bulge are to be preferred, although progenitors without a bulge and
different mass-ratios appear to be favored when the tilt $\lambda$ is
estimated. However the rather small statistics of our study precludes a
conclusive answer.

%%%%%%%%%%%%%%%%%%%%%%%%%%%%%%%%%%%%%%%%%%%%%%
\subsection{Luminosity - Phase Space Relation}
%%%%%%%%%%%%%%%%%%%%%%%%%%%%%%%%%%%%%%%%%%%%%%

 The {\it effective} coarse-grained 
phase-space density of our numerical remnants is computed using equation (2.12)
of HSH  for the luminous matter,  namely:
\begin{equation}
f_{\rm e} \equiv  \frac{1}{\sigma_0\, R_{\rm e}^2} \,,
\label{eq:phase}
\end{equation}
where the gravitational
constant $G$ has been taken to be unity.   This
estimator allows a more direct comparison of the global trend of the
coarse-grained central phase-space
with $M_K$ of our numerical remnants with their
observational counterparts. 
  A direct least-squares fit of the form 
\begin{equation}
\log L_K  \propto - \gamma \log f_{\rm e} \;,
\label{eq:magtotal}
\end{equation}
where  $\gamma$ is the fitting parameter, is done to
the mean of the values of the projections of each remnant. The magnitude value
is taken as $M_K = -2.5\log L_K$. A similar fit
  is done to the observational data of MGAZ99.

\begin{figure}[!t]
\centering
\includegraphics[width=8.5cm]{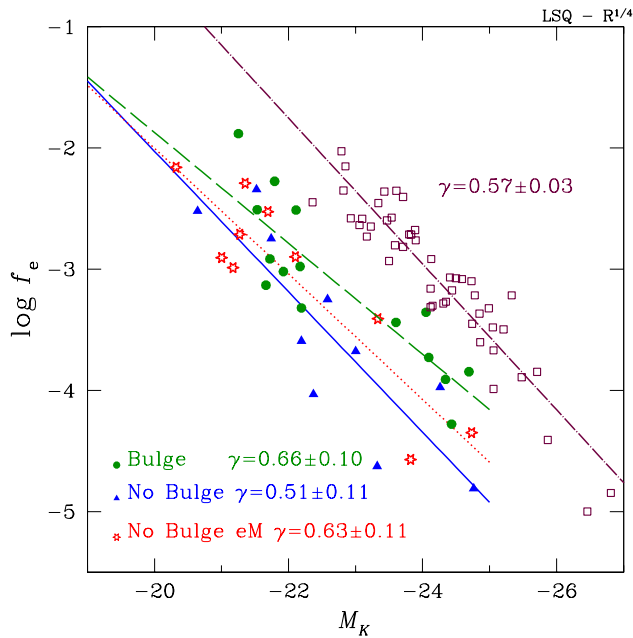}
\caption{Effective coarse-grained phase-space density $f_{\rm e}$ 
versus total magnitude $M_K$.  
 The $K$-band data sample of ellipticals from MGAZ99 ({\it open-squares,
   dot-dashed line})
 yields a scaling index of $\gamma=0.57$.  The scalings for our remnants with
 a bulge ({\it  long-dashed, solid circles}) and without it ({\it  solid line
   and triangles}) are shown; the bulge-less equal mass merger are also plotted
 ({\it dotted line and  open stars}). 
    } 
\label{fig:phase}
\end{figure}

The correlations $M_K$--$f_{\rm e}$ for our numerical remnants corresponding to
progenitors with  
({\it solid dots}) and without a bulge
component  ({\it solid triangles}) are indicated in
Figure~\ref{fig:phase}.  Results for the equal-mass
  mergers are shown with an open-star symbol.
For comparison, 
the $M_K$--$f_{\rm e}$ relation derived from the data of MGAZ99 is also
shown ({\it open  squares}). The index $\gamma$ obtained for the different
cases considered in this work are shown in Table~3.

A scaling relation of the form  $L\propto f_{\rm
  e}^{-\gamma}$ results from these fits, with $\gamma=0.66$ and  $\gamma=0.51$
  for our remnants  (using a $R^{1/4}$ profile and {\sc lsq}) from progenitors
  with and   without a bulge, respectively. For equal-mass mergers, we
  have $\gamma=0.63$.

The value 
  for the MGZA99 sample is $\gamma=0.57$.  
 This value is similar to that obtained in the
  B-band for the data of Bender et~al. (1992) by HSH ($\gamma=0.54$), and that
  if we use the
  ellipticals from D'Onofrio (2001) ($\gamma=0.51$); despite the fact
  that D'Onofrio   uses a  S\'ersic profile  to fit  the
  brightness distribution while  Bender et~al. consider a $R^{1/4}$
  profile.

In general good agreement, within uncertainties, among the $N$-body
  and observed $\gamma$ indices exist. 
  Taken at face value,  non-equal mass
  (bulgeless) mergers have a closer value to the observed
  ones. However, as for the FP case, our
  results preclude us from favoring a particular model for the
  progenitors.

%%%%%%%%%%%%%
%%%%%%%%%%%%

%%%%%%%%%%%%%%%%%%%%%%%%%%%%%%%%%%%%%%%%%%%%%%%%%%%%%%%%%%%%%%%%%%%%%%%%%
%%%%%%%%%%%%%%%%%%%%%%%%%%%%%%%%%%%%%%%%%%%%%%%%%%%%%%%%%%%%%%%%%%%%%%%%%
\section{Discussion and Final Comments}
%%%%%%%%%%%%%%%%%%%%%%%%%%%%%%%%%%%%%%%%%%%%%%%%%%%%%%%%%%%%%%%%%%%%%%%%%
%%%%%%%%%%%%%%%%%%%%%%%%%%%%%%%%%%%%%%%%%%%%%%%%%%%%%%%%%%%%%%%%%%%%%%%%%

An important conclusion of recent works on mergers (e.g. R06, Naab \& Trujillo
2006, Cox et al. 2006,
Dekel \& Cox 2006) is that dissipation plays a major role in reproducing
different scaling relations; such as the FP or the $R_{\rm e}$--$M_*$
relation.  In particular, R06 find that the initial fraction of gas in merging
galaxies should be $\approx 30$\% of the disk mass in order to reproduce both
the observed scaling of the  
Fundamental Plane and the $R_{\rm e}$--$M_*$ relation.

However, previous works by AV05 and AVC06 were able to reproduce rather
adequately  
these global scaling relations without invoking a gas component or
even a bulge-like structure. The $b$ index of the FP was less well
reproduced, perhaps due to the assumption of a constant $\Upsilon_*$
to obtain $\langle I \rangle_{\rm e}$. The results found here for 
  progenitors  with a bulge, and bulge-less equal-mass mergers, are in 
  general in same standing when compared with NIR data under
  methods that mimic the observational procedure.

As mentioned earlier, the observed tilt $\lambda$ is  well
  reproduced in our   dissipationless simulations. The reason for this
  agreement, although the $b$   index is rather poorly matched
  (see Table 3), might be
  related to  correlations present  among the variables (and their errors)
  involved in the fitting of the  FP. Such correlations will
  affect the fitting parameters depending on how the FP is fitted, either as
  in equation (\ref{eq:fplane}) or in (\ref{eq:fplane2}).

On the other hand, it is not clear the origin of the
difference  between our  fitted values of
$a$ and $b$  of the FP and those derived by 
Robertson et al. (2006) for their equal-mass  
    dissipationless simulations.  For
    instance, when we 
    use the half-projected stellar mass radius, $R_{\rm 50}$,
    and the $1$-D 
    velocity dispersion, $\sigma_{50}$, inside this radius we obtain a
    value of $a\approx 1.8$ that tends to the expected virial
    value as in R06. However, even in this case our value for $b\approx 0.5$
    is far from unity (see Table~3). 
   The tilt $\lambda\approx 0.8$ in this case tends to increase a
   little but still it is not unity.

Given our results, the FP appears to arise mainly from broken
structural and dynamical homology between the observed (luminous component)
quantities $\{\sigma_0, R_{\rm e},\langle I\rangle_{\rm e}\}$ and their
theoretical counterparts $\{V^2,R_{\rm G},M\}$; where $V^2=2T/M$, $R_{\rm
  G}=GM^2/|U|$, and $M$ is the total bound mass of the system. Here 
 $T$ is the total kinetic energy of the system, with contributions
from rotational and thermal motion in remnants. An indication of the
previous was presented in Aceves \& Vel\'azquez (2005b). 
Variations in the stellar mass-to-light ratio, $\Upsilon_*$, can be 
responsible for some of the tilt in the FP, especially if we focus on  the
value of the $b$ index for our remnants. However, the effect of $\Upsilon_*$
would tend, according to our results, to diminish when the $\lambda$ parameter
is  used as an indicator of the FP tilt.

The $L$--$f_{\rm e}$ scalings shown in Figure~2 are in the same direction
 as the preceding results on the FP,
 suggesting that  the establishment of this scaling relation is consistent
 with a dissipationless merging scenario. It is likely that
 dissipative processes (e.g. star formation and feedback) will play a
 major role in defining the zero-point of all 
 these scaling relations, and affect the indices in a not simple way.

From  energy
arguments in a dissipationless scenario, and
    assuming equal mass  galaxies, Hernquist et al.~(1993) derived the
    particular relation  $L \propto f_{\rm e}^{-0.5}$. When
    non-identical progenitors 
are taken into account these energy arguments predict a shallower 
log-slope; for example, for a mass ratio of about $1/5$ it leads to
$L \propto f_{\rm  e}^{-0.37}$.  So, according to these energy
arguments we should expect a range of values in the interval $0.37
\lesssim \gamma \lesssim 0.5$ for the  non-equal mass ratios considered
here.  Thus, the
    theoretical expectation on the basis of these energy arguments is about
    $30$\% lower than the average value ($\langle \gamma\rangle \approx 0.7$)
    for our no-bulge 
    equal-mass mergers. Meanwhile these theoretical values are more
    closer to the ones of our remnants from bulgeless progenitors.

The average result from all of our mergers yield an  index
  $\langle \gamma \rangle \approx 0.65$ that is in good  agreement with the
observed NIR value 
  $\gamma \approx 0.60$. This agreement spans for
about four magnitudes and compromises remnants that can be cataloged
from bright dwarf ellipticals to intermediate early-type
galaxies (see Figure~2).

It should be noticed that
differences arise when a primordial bulge is
  considered or not, and merger remnants from bulgeless
  progenitors tend to produce a scaling more close to the NIR one. In
  general, for larger ``luminosity'' the 
  effective  phase-space density tends to a higher value for remnants with
  progenitors  having a bulge as is found in other works that quantify
  the physical distribution function $f_{\rm p}$ (e.g. HSH,
  Naab \& Trujillo 2006, Robertson et al. 2006). At  low
  ``luminosity'' the behavior of $f_{\rm e}$ does not show a clear
  difference  for bulge or bulgeless progenitors. 

A physical explanation for the details of the behavior of $L$--$f_{\rm
  e}$ (or it physical counterpart) is non-existent at the moment,
  although the arguments of HSH yield light on the subject. 
On other hand, unfortunately,  the
  more extended  study carried out by R06 did not address the $L$--$f_{\rm 
    e}$  scaling relation to compare directly with our $N$-body results.

 In summary, the observed $\lambda$-tilt of the FP and the $\gamma$--index
  of the luminosity-phase space relation, both in the NIR, are consistent
  with those obtained here through dissipationless merger simulations; at
  least over the luminous mass range covered by our remnants.  We
  ascribe such concordance to the use of progenitors constructed under a
cosmological motivated model of galaxy formation, with a 
NFW dark halo and satisfying initially a Tully-Fisher relation.

 As a concluding remark,  given the results of this work, we would like
 to point out  that 
trying to deduce the amount of primordial gas ($f_{\rm gas}$) in
models of progenitors,  taking
  part in merger events, to satisfy a global scaling relation like the
FP,  might be subject to a non-negligible uncertainty. It is likely that the
zero-point of the scaling relations would be a better constraint to
the dissipational component of disk galaxy progenitors. In our opinion this
is a subject that requires  further work, but it is out of the scope of the
present work.

\acknowledgments
%%%%%%%%%%%%%%%%%%%%%%%%%%%
 
This research was funded by UNAM-PAPIIT Research Project IN121406, and 
  CONACyT Project 25030. The simulations presented here were run on
{\sc Kan Balam} of
  DGSCA-Departamento de Superc\'omputo of the Universidad Nacional Aut\'onoma
  de M\'exico, and in the Centro Nacional de Superc\'omputo de San Luis
  Potos\'{\i}.

%%%%%%%%%%%%%%%%%%%%%%%%%%%

%%%%%%%%%%%%%%%%%%%%%%%

\end{document}